\documentclass[twocolumn]{aastex62}
\usepackage[]{amsmath}

\graphicspath{{./}{figs/}}

\received{-}
\revised{-}
\accepted{-}
\submitjournal{ApJ}

\shorttitle{}
\shortauthors{Teague et al.}

\begin{document}

\title{Evidence For A Vertical Dependence on the Pressure Structure in AS~209}

\correspondingauthor{Richard Teague}
\email{rteague@umich.edu}

\author[0000-0002-0786-7307]{Richard Teague}
\affil{Department of Astronomy, University of Michigan, 311 West Hall, 1085 S. University Ave, Ann Arbor, MI 48109, USA}

\author[0000-0001-7258-770X]{Jaehan Bae}
\affil{Department of Terrestrial Magnetism, Carnegie Institution for Science, 5241 Broad Branch Road, NW, Washington, DC 20015, USA}

\author[0000-0002-1899-8783]{Tilman Birnstiel}
\affil{University Observatory, Faculty of Physics, Ludwig-Maximilians-Universit\"{a}t M\"{u}nchen, Scheinerstr. 1, D-81679 Munich, Germany}

\author[0000-0003-4179-6394]{Edwin A. Bergin}
\affil{Department of Astronomy, University of Michigan, 311 West Hall, 1085 S. University Ave, Ann Arbor, MI 48109, USA}

\begin{abstract}
We present an improved method to measure the rotation curves for disks with non-axisymmetric brightness profiles initially published in \citet{Teague_ea_2018a}. Application of this method to the well studied AS~209 system shows substantial deviations from Keplerian rotation of up to $\pm 5\%$. These deviations are most likely due to perturbations in the gas pressure profile, including a perturbation located at $\approx 250$~au and spanning up to $\approx 50$~au which is only detected kinematically. Modelling the required temperature and density profiles required to recover the observed rotation curve we demonstrate that the rings observed in $\micron$ scattered light are coincident with the pressure maxima, and are radially offset from the rings observed in mm continuum emission. This suggests that if rings in the NIR are due to sub-$\micron$ grains trapped in pressure maxima that there is a vertical dependence on the radius of the pressure minima.
\end{abstract}


\section{Introduction}
\label{sec:intro}

Long baseline observations with the Atacama Large (sub-)Millimetre Array (ALMA) have shown that substructures in the thermal continuum of protoplanetary disks are likely ubiquitous. These features are frequently interpreted in the context of gas pressure maxima into which grains are shepherded through complex gas-grain interactions \citep{Birnstiel_ea_2012, Pinilla_ea_2012}.

Identification of the main driver of these pressure maxima, such as an unseen planet or (magneto-) hydrodynamical instabilities, is hampered by the lack of a reliable tracer of the gas pressure profile. Although CO isotopologues are routinely found to exhibit structure in their emission profiles \citep{Isella_ea_2016, Fedele_ea_2017}, relating these to an accurate surface density profile requires several assumptions about the local physical and chemical conditions to be made.

Recently, \citet{Teague_ea_2018a} demonstrated a technique to measure highly precise rotation velocities in axisymmetric disks. For a geometrically thick disk with gradients in temperature and density the rotation velocity is given by,

\begin{equation}
\frac{v_{\rm rot}^2}{r} = \frac{GM_{\star} r}{(r^2 + z^2)^{3/2}} + \frac{1}{\rho_{\rm gas}} \frac{\partial P}{\partial r},
\label{eq:rotation}
\end{equation}

\noindent where $P = n_{\rm gas}kT$ is the gas pressure and $\rho_{\rm gas}$ is the gas density. Thus, in combination with measurements of the emission surface \citep{Pinte_ea_2018a}, deviations from Keplerian rotation can be used to infer the pressure gradient. \citet{Teague_ea_2018a} used this to place tight constraints on the gas surface density profile of HD~163296 and infer the presence of two Jupiter-mass planets.

Others have also advocated the use of kinematics to identify potential sources for changes in the pressure gradient. \citet{Pinte_ea_2018a} reported kinematic evidence of a wide separation $\sim 2 M_{\rm Jup}$ planet at $\approx 260~{\rm au}$ in HD~163296, extending far beyond the continuum edge. Similarly, \citet{Perez_ea_2018} showed that planet-disk interactions will drive large non-Keplerian velocities which can be used to locate potential perturbers.

In addition to searching for the signs of embedded protoplanets, constraints of the pressure gradient are invaluable for interpreting observations of the dust. The inward radial motion of particles due to the headwind from gas rotating at sub-Keplerian speeds, as the gas is supported by the radial pressure gradient, can very rapidly deplete the disk of dust. To slow this depletion, pressure bumps are frequently invoked, resulting in the trapping of particles and thus extending the lifetime of the dust disk \citep{Pinilla_ea_2012}. However, despite the necessity of such pressure traps, direct evidence of changes in gas pressure (rather than local enhancements of dust interpreted as a dust trap) are lacking.

In this paper we present an improved method to measure the rotational velocity of a protoplanetary disk which relaxes assumptions of the azimuthal symmetry and intrinsic Gaussian line profiles which is described in Section~\ref{sec:methodology}. In Section~\ref{sec:as209}, we apply this method to archival ALMA data of AS~209 and present a discussion of the observed features. We summarise the findings and conclude in Section~\ref{sec:discussion}.

\section{Measuring The Rotation Velocity}
\label{sec:methodology}

In \citet{Teague_ea_2018a} we presented a method to measure the rotation velocity of an axisymmetric disk. The method required the minimization of the width of the averaged line profile at a given radius after accounting for the projected rotation. While this method proved to be robust, it makes the assumption that the resulting averaged profile would be a single Gaussian component. We have improved upon this technique to relax this assumption by modelling the stacked spectrum as a Gaussian Process which allows a much more flexible model \citep{Foreman-Mackey_ea_2017}. In this section we review the method and describe the updates.

\begin{figure*}
\centering
\includegraphics[]{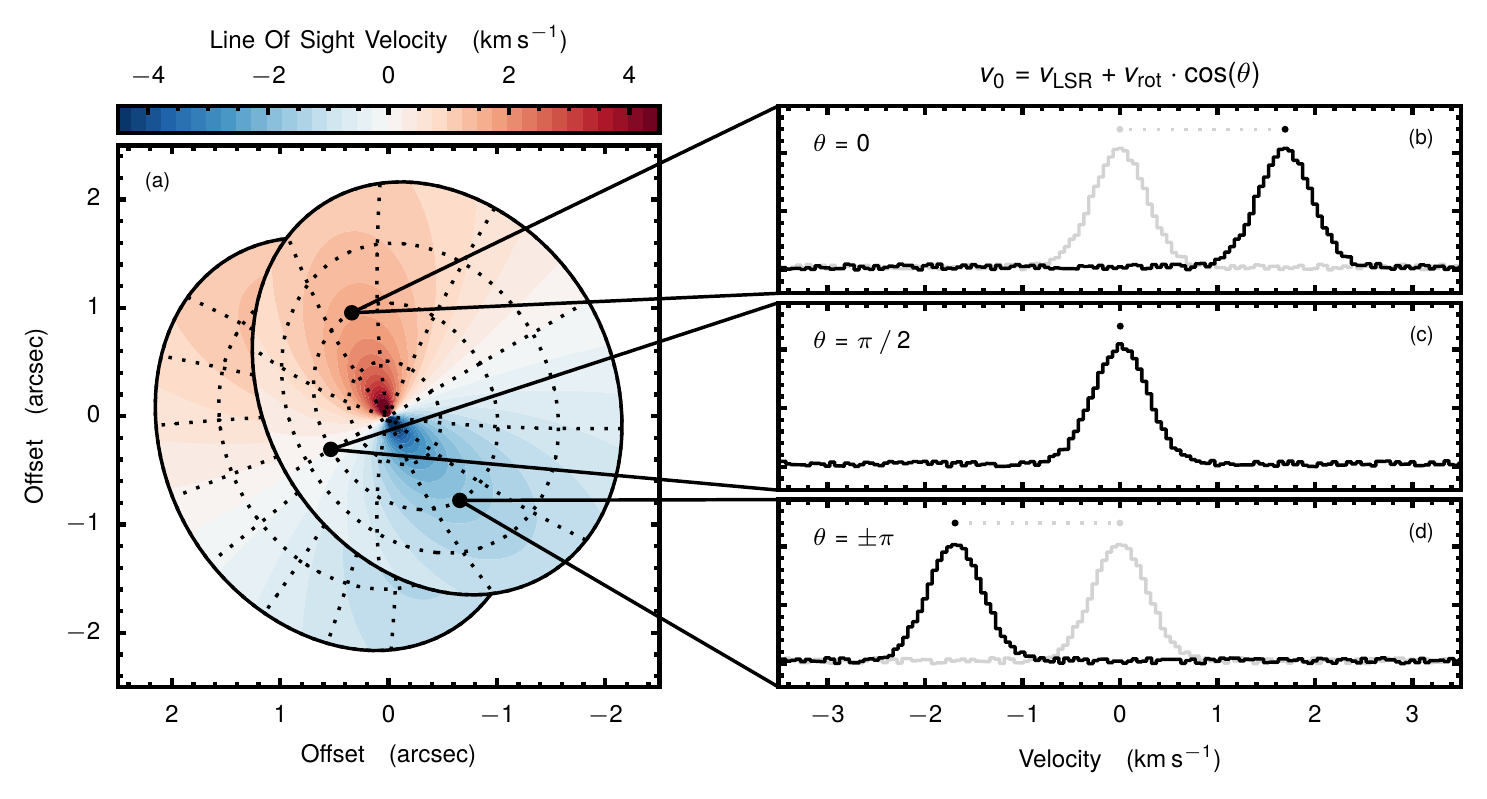}
\caption{Demonstration of the method with a toy model. Panel (a) shows an example for the projected velocity arising from a flared disk structure. The dotted lines show lines of constant radius and polar angle. Panels (b), (c) and (d) show spectra extracted at the marked locations in black. The gray lines show the same spectra but shifted by $-v_{\rm rot} \cdot \cos(\theta)$, back to the systemic velocity of the disk and thus able to be stacked. \label{fig:first_moment_and_spectra}}
\end{figure*}

We make use of the fact that protoplanetary disks are predominantly azimuthally symmetric. Line emission arising from the same radial location in the disk should therefore be tracing the same physical and chemical properties and thus possess the same profile. The only difference will be in the line center which will be offset from the systemic velocity\ by $v_{\rm rot} \cdot \cos (\theta)$ where $\theta$ is the polar angle measured from the red-shifted major axis. Note that this is not the polar angle measured in the sky plane but must be calculated taking into account the disk geometry (see the radial dotted lines in Fig.~\ref{fig:first_moment_and_spectra}).

Figure~\ref{fig:first_moment_and_spectra} shows a toy model as an example. In panel (a) the projected line-of-sight velocity is shown by the filled contours. As the disk is flared, we see both the top side and far side of the disk and this demonstrates the need to correctly account for the emission height of the disk. The dotted lines trace lines of constant $r$ and $\theta$. Spectra extracted at the black dots, as shown in the panels (b), (c) and (d), will be the same shape, but offset on the velocity axis (shown in black). If $v_{\rm rot}$ is known, these spectra can be shifted back to the systemic velocity, ready to be stacked, as shown by the gray lines. This technique has been used previously to significantly boost the SNR of spectra and increase sensitivity in the outer edge of the disk \citep{Teague_ea_2016, Yen_ea_2016, Matra_ea_2017} and a similar method used by \citet{Yen_ea_2018} to measure dynamic stellar masses.

Rather than assuming $v_{\rm rot}$ \emph{a priori} we can use the deprojected spectra to infer what the correct value is. In \citet{Teague_ea_2018a} we used the linewidth of the stacked spectrum as a proxy. If the lines are deprojected using an incorrect value, the line centers will still have some offset leading to a broadening of the final line profile. Thus, the $v_{\rm rot}$ value which minimizes the final linewidth is the correct value as this is when all the spectra are correctly aligned. However, this assumes that the final profile is Gaussian which is often not the case. For example, optically thick lines, such as $^{12}$CO and potentially $^{13}$CO, will have line profiles which deviate significantly from a Gaussian due to the saturation of the line core.

Here we argue that a more flexible approach is to model the stacked spectrum as a Gaussian Process (essentially a probabilistic approach to modelling smooth, non-parametric functions) and find the value of $v_{\rm rot}$ which minimizes the variance in the residuals. A similar approach has been used by \citet{Czekala_ea_2017} to model the spectra of binaries.

\begin{figure*}
\centering
\includegraphics[]{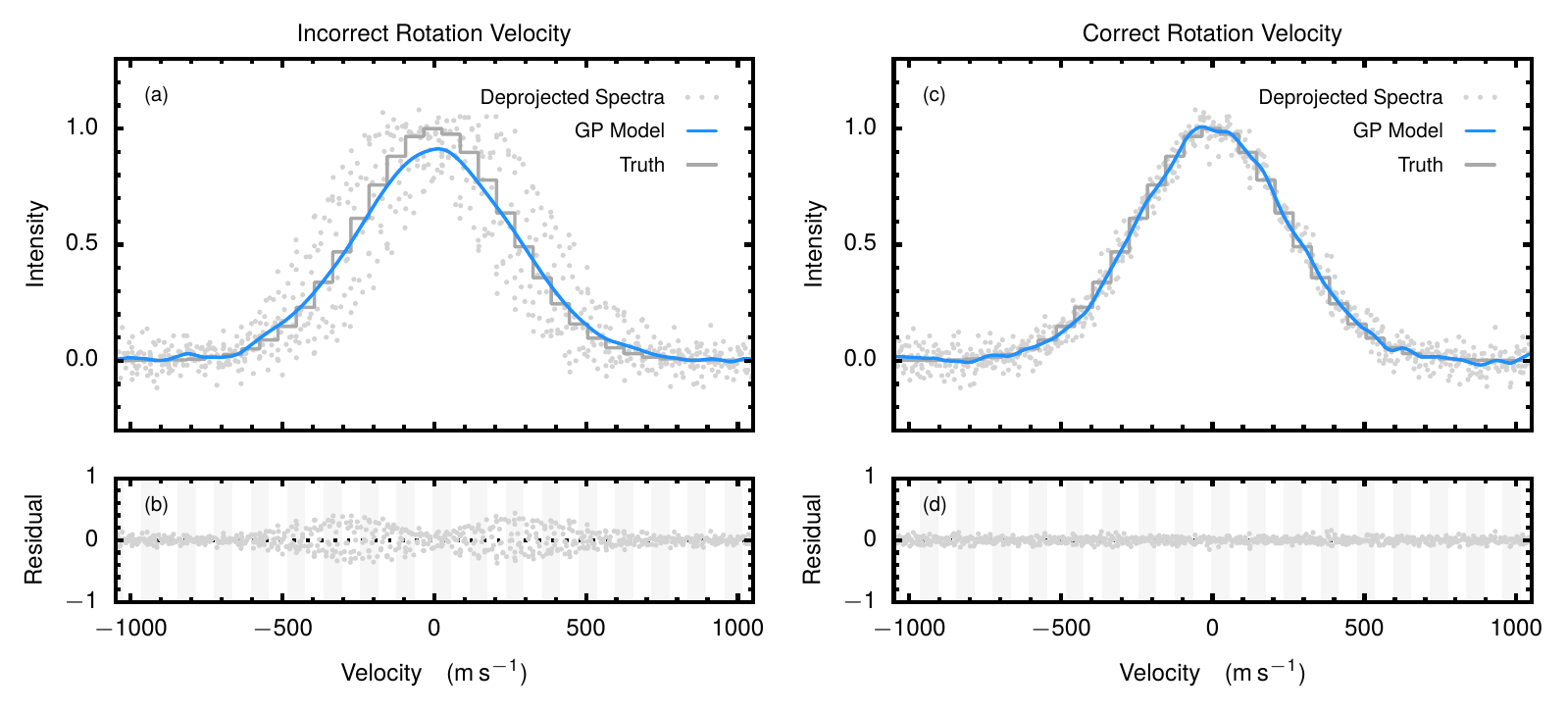}
\caption{Demonstration of the quality of fit method used. Each point represented a channel in a spectrum given some velocity shift. The blue lines are the Gaussian Process (GP) model of these data and the gray solid line shows the true intrinsic line profile. The bottom row shows the residual between the individual points and the GP model. If an incorrect $v_{\rm rot}$ is used to deproject the data (left column) significant variance is seen in the residuals. Conversely, when the correct $v_{\rm rot}$ is used (right column), the variances in the residuals is minimised. The shading in the residual plots shows the channel width, demonstrating that after the deprojection of all the spectra, we sample the intrinsic line profile at a much higher rate than the observations, allowing us to achieve a much higher precision on $v_{\rm rot}$ than the native channel width. \label{fig:deprojection_example}}
\end{figure*}

An example of this approach is shown in Figure~\ref{fig:deprojection_example} where the left column shows the situation where an incorrect $v_{\rm rot}$ value has been assumed for the deprojection and thus there is significant scatter in the residual around the line wings, as shown in panel (b). Conversely, when the correct $v_{\rm rot}$ has been used, as shown in the right column, the scatter in residuals is near constant across the profile. This allows for the case of highly non-Gaussian profiles or if there is significant differences in line brightness as a function of azimuth.

Figure~\ref{fig:deprojection_example} also demonstrates why a precision well below the velocity resolution can be achieved. As the line profile is sampled at different locations due to shifts from the rotation, the deprojected spectra result in a sampling of the intrinsic line profile at a factor of $\sim 10$ higher. The gray bars in the residual plots show the width of a single channel and demonstrate that the profile is sampled at a much higher frequency after the shift. Although some broadening may be present due to the Hanning smoothing applied in the correlator, the resulting $v_{\rm rot}$ will be insensitive to this as the optimization does not care about the properties of the stacked line profile other than it is smooth.

In practice this is performed using the Python package \texttt{celerite} \citep{Foreman-Mackey_ea_2017}. This approach has the added advantage that a more robust uncertainty can be derived for $v_{\rm rot}$. As \texttt{celerite} naturally considers correlations between pixels, any spatial correlations arising from the imaging \citep[such as those described in][]{Flaherty_ea_2018} will be accounted for in the uncertainty. Furthermore, this approach more readily allows for the inclusion of priors for $v_{\rm rot}$, or a more specific model for the noise \citep[for example the correlated noise described in][]{Teague_ea_2018b}.

We have tested this method with various levels of noise, velocity resolutions, azimuthal asymmetries in both line width and peak and non-Gaussian lines profile. A thorough examination of the accuracy and precision achieved is presented in Appendix~\ref{sec:app:rotation_velocity}, including the impact of azimuthal structure in the line profiles or when the near and far sides of the disk are spatially resolved. In brief, however, the accuracy achieved by this method when there is no strong azimuthal structure can be estimated by

\begin{equation}
44.4 \pm 0.1~{\rm m\,s^{-1}} \times \left( 
\sqrt{\frac{2 \pi r}{\theta_{\rm beam}}} \cdot \frac{{\rm SNR}}{10} \right)^{-1.12 \pm 0.01}
\end{equation}

\noindent where SNR is the signal-to-noise achieved in a single pixel at radius $r$ and $\theta_{\rm beam}$ is the beam FWHM. Azimuthal deviations in the linewidth and peak do not significantly hinder this approach unless they are greater than $\approx 50\%$ in magnitude.

Examples of the code used for this and Jupyter Notebooks containing guides on how to use them can be found at \url{https://github.com/richteague/eddy}. Version 1.0 of the code used for this paper can be obtained from \url{https://doi.org/10.5281/zenodo.1440051}.

\section{The Rotation Curve of AS~209}
\label{sec:as209}

To demonstrate this method we use archival data of AS~209 (2015.1.00486.S, PI Fedele, D.). Continuum observations from this project have been previously presented in \citet{Fedele_ea_2018} which have suggested that multiple gaps in the continuum can be driven by a single planet. Here we focus only on $^{12}$CO emission, leaving a thorough analysis of the three CO isotopologues and DCO$^+$ to be presented in a future work (Favre et al., in prep.).

\subsection{Observations}
\label{sec:as209:observations}

\begin{figure*}
\includegraphics[width=\textwidth]{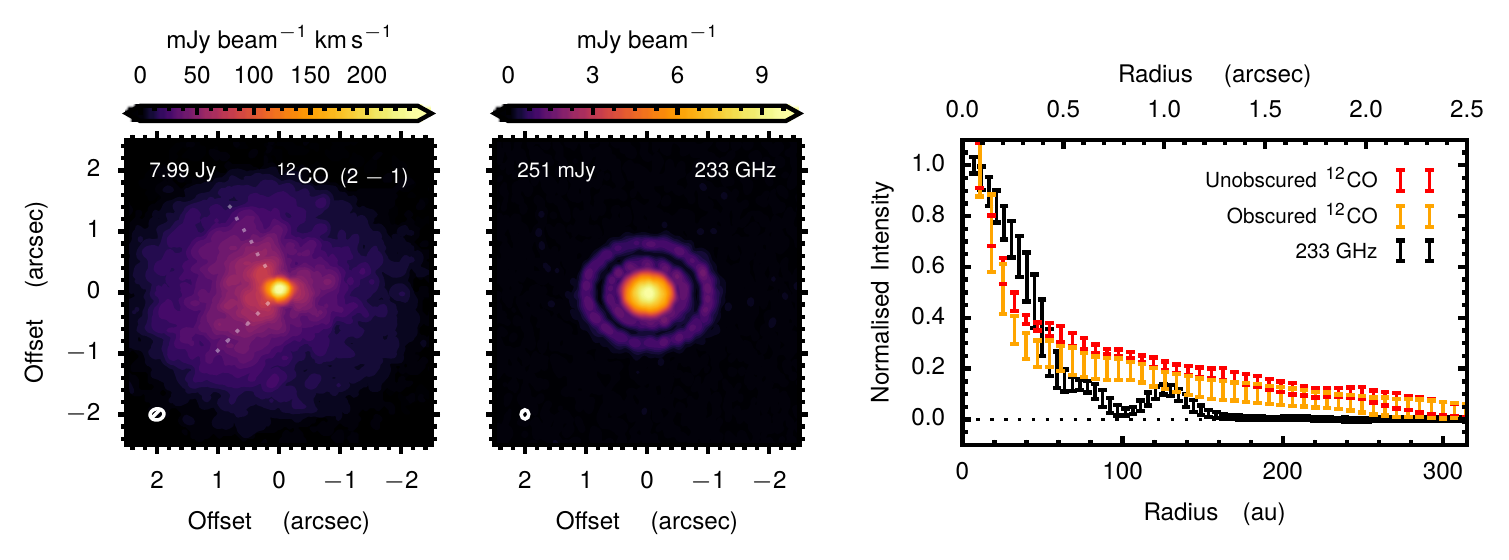}
\caption{Observations of $^{12}$CO and the 233~GHz continuum. The left two panels show the integrated intensities while the right panel shows the normalised azimuthally averaged profile. Synthesized beamsizes are shown in the bottom left of each panel. The integrated flux is shown in the top left. The coloured contours have been saturated in the centre to highlight the extended emission. The radial profiles are sampled at quarter beam spacing and the error bars show the standard deviation of the annulus. $^{12}$CO emission is split into `unobscured' and `obscured', east and west of the dotted lines (at $\pm 55\degr$ either side of the major axis) shown in the left most panel. \label{fig:observations}}
\end{figure*}

Data reduction followed the same process as outlined in \citet{Fedele_ea_2018}. The data were calibrated using the provided scripts in \texttt{casa v4.4} before moving to \texttt{casa v5.2} for the imaging and self-calibration. Phase gain tables were calculated on the continuum window then applied to the three spectral line windows containing $^{12}$CO emission.

We consider two cases, both with and without the continuum subtracted from the line data. The continuum is removed using the task \texttt{uvcontsub} which linearly interpolates the the continuum from line-free channels in the $uv$-plane. As discussed in \citet{Boehler_ea_2017}, this can lead to an under-estimation of the true total intensity of the line as the molecular gas will absorb some of the continuum. While this will affect the inferred temperature or column density which require absolute flux measures, our method is insensitive to such effects. As the effect is essentially independent of frequency, at least across the line profile, see for example Fig.~8 in \citet{Boehler_ea_2017}, then the change in the line profile will be symmetric about the line center and thus not change our derivation of velocity.

Imaging the continuum emission we used uniform weighting resulting in a beamsize of $0.15\arcsec \times 0.13\arcsec$ at a position angle of $2.6\degr$. A RMS noise of $\sigma = 65~\mu {\rm Jy~beam^{-1}}$ was measured in continuum free regions of the continuum map and an integrated intensity of 251~mJy was measured, consistent with previous observations \citep{Oberg_ea_2011, Huang_ea_2016}.

We perform a different approach for the $^{12}$CO emission. Using the \texttt{tclean} task, we first image the emission with natural weighting, to maximize sensitivity, and with a square $3\arcsec \times 3\arcsec$ box as the mask. From this image we generate a first moment map clipping values below $2\sigma$ where $\sigma$ was measured in a line free channel. To this first moment map we fit a Keplerian profile to derive a position angle, $v_{\rm LSR}$ and $M_{\rm star}$, holding the inclination constant at $i = 35.3\degr$ \citep{Fedele_ea_2018} in order to break the $M \cdot \sin i$ degeneracy. These values were then used to generate a Keplerian mask (masking out regions where, given the derived rotation pattern, we would not expect emission to arise) for the data which was convolved with the beam and checked to encompass the whole emission. The data were then imaged and CLEANed again using this Keplerian mask.

For precise measurements of $v_{\rm rot}$, sensitivity is more important than spatial resolution. Therefore we use natural weighting for the imaging yielding beam sizes of $0.23\arcsec \times 0.19\arcsec$ at a position angle of $-79.3\degr$. The data were imaged at a velocity resolution of $\approx 160~{\rm m\,s^{-1}}$ ($\approx 244~{\rm kHz}$). The RMS noise in a line free channel was $3.3~{\rm mJy~beam^{-1}}$ and we measure an integrated flux of 7.99~Jy~km~s$^{-1}$, consistent with previous measurements \citep{Huang_ea_2016}.

Figure~\ref{fig:observations} summarises the observations, showing the velocity integrated fluxes (using a $2\sigma$ clip and the CLEAN mask), and their radial profiles. Significant absorption is seen in the west half of the $^{12}$CO emission, likely due to cloud contamination at $v_{\rm LSR} \lesssim 5~{\rm km~s^{-1}}$ \citep{Oberg_ea_2011, Huang_ea_2016}. Assuming that for $v_{\rm LSR} \gtrsim 5~{\rm km~s^{-1}}$ the emission is free from absorption, ratios of the intensity profiles suggest that the cloud absorbs $\approx 30\%$ of the $^{12}$CO emission on the Western side of the disk.

There are no clear deviations from a Keplerian pattern observed in the channel maps indicative of large scale kinematic features \citep{Perez_ea_2015, Pinte_ea_2018b}, however the cloud contamination and limited sensitivity may limit the visibility of such features.

\subsection{Measuring Rotation Curves}
\label{sec:as209:rotation_curves}

In order to use the method presented in Section~\ref{sec:methodology} an initial estimate of the rotation profile and the emission height, in order to properly deproject the data into annuli of constant radii are required.

For an initial estimate of the expected rotation profile we fit a Keplerian pattern, including a conical emission surface \citep{Rosenfeld_ea_2013} using the MCMC ensemble sampler \texttt{emcee} \citep{Foreman-Mackey_ea_2013}. We calculate a map of the line-of-sight velocities using the method presented in \citet{Teague_Foreman-Mackey_2018} which fits a parabola to the pixel of peak intensity and its two neighbouring pixels. This method allow us to discriminate between emission arising from the near side of the disk and the far side, while achieving a sub-channel precision measurement of the line centroid. In addition, the cloud absorption will less strongly bias the measurement of the maximum coordinate. The emission surface is calculated as $z = r \cdot \tan \psi$ where $\psi$ is the angle between the disk midplane and the emission surface\footnote{We have also tried a more complex surface of $z = z_0 \times r^{\phi}$, however the spatial resolution of the data meant that the fits were unable to converge.}. In addition, with each call of the likelihood function the rotation pattern was convolved with a 2D Gaussian matching the synthesized beam for each observation to account for convolution effects in the inner regions of the disk \citep{Walsh_ea_2017}.

The best-fit rotation pattern, assuming a fixed inclination $i = 35.3\degr$, was described with $M_{\star} = 1.16 \pm 0.01~M_{\rm sun}$, ${\rm PA} = 86.7\degr \pm 0.1\degr$,  $v_{\rm LSR} = 4670 \pm 5~{\rm m\,s^{-1}}$ and $\psi = 13.1\degr \pm 0.3\degr$. The uncertainties are the standard deviation of the posterior distribution. The inferred stellar mass is slightly larger than the previously found $M_{\rm star} = 0.9~M_{\rm sun}$ \citep{Andrews_ea_2009}, likely due to better resolving the emission in high velocity channels which contain the most information to distinguish between stellar masses. The position angle and systemic velocity are consistent with previous determinations \citep{Andrews_ea_2009, Huang_ea_2016}.

Using the method presented in \citet{Pinte_ea_2018a} to measure the emission height we find good agreement with a conical model with $\psi \approx 13\degr$, consistent with the determination from the ninth moment map fitting. As the emission surface determination relies on the asymmetry across the major axis of the disk, cloud absorption will not impact this result as this only results in an asymmetry across the minor side of the disk. Offsets in the ellipse centres were found in the northern direction meaning that the southern side of the disk is closer consistent with the preferred orientation proposed by \citet{Avenhaus_ea_2018} who showed that features in the NIR scattered light better align with continuum features when the southern side, rather than the northern side is closer. This orientation is in addition consistent with that found from the fitting of the ninth moment map.
 
\begin{figure*}
\centering
\includegraphics[]{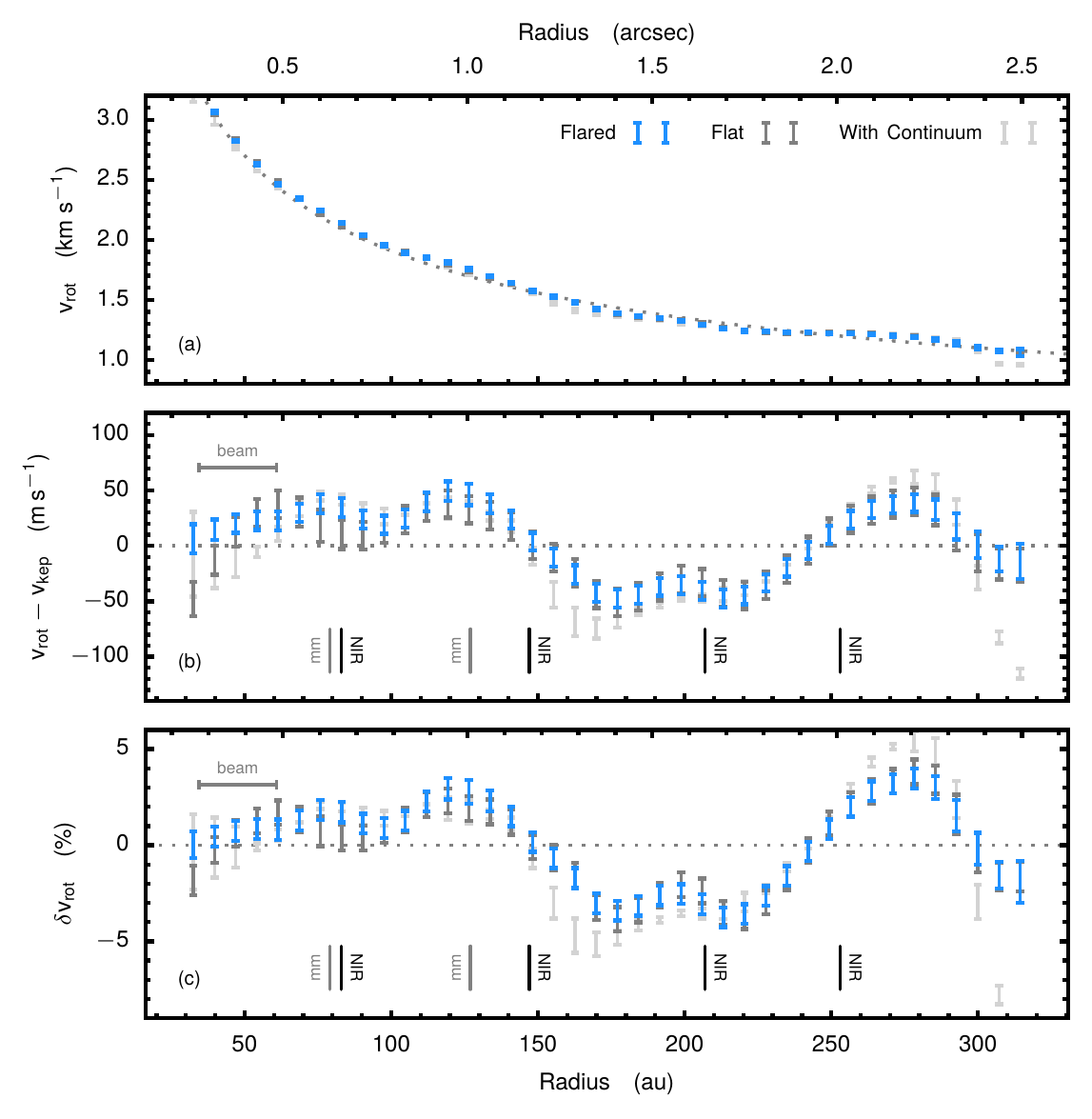}
\caption{Residuals between the measured $v_{\rm rot}$ and the Keplerian rotation curve found from fitting the first moment map. Error bars show $1\sigma$ uncertainties with blue assuming a flared surface for the deprojection, dark gray bars assuming a geometrically thin disk and light gray bars a flared disk without continuum subtraction. The vertical lines show the centre of the rings in dust mm emission \citep[gray,][]{Fedele_ea_2018} and scattered light \citep[black, assuming that the southern side is closer to the observer,][]{Avenhaus_ea_2018}. The synthesised beamsize is shown in the top left corner. \label{fig:rotation_residual}}
\end{figure*}

Following the method in \citet{Teague_ea_2018a} and with the modification described in Section~\ref{sec:methodology} we measure an azimuthally averaged rotation profile. We consider three cases: firstly we deproject the sky-plane coordinates into disk coordinates to correctly account for the flaring, secondly we consider a geometrically thin disk, and finally we consider the flared disk with no continuum subtraction. For each we we then bin the data into annuli with a width of a quarter of the beam ($\approx 0.05\arcsec$) and derive a $v_{\rm rot}$ value. We further model the radial profile as a Gaussian Process, requiring the profile to be smoothly varying and to account for the correlations due to the sub-beam size sampling.

These rotation profiles are shown in Figure~\ref{fig:rotation_residual}a, with $1\sigma$ uncertainties. The dotted line is a fit of a geometrically thin Keplerian rotation profile assuming $i = 35.7\degr$ and resulting in $M_{\rm star} = 1.25~M_{\rm sun}$. This profile is used as the reference profile for the derived values, however we stress that this should not be taken at the true stellar mass as we are unable to disentangle large scale effects of the pressure gradient from the stellar mass. The middle and bottom panels show the residual in ${\rm m\,s^{-1}}$ and as a percentage, respectively. Gray and black lines show the locations of rings observed in the mm continuum and NIR scattered light, respectively. 

All three scenarios yield comparable radial $v_{\rm rot}$ profiles which are consistent within $3\sigma$ of one another. This suggests that continuum subtraction does not significantly affect the derived rotation profile.

When plotting residuals from a Keplerian profile, as in panels (b) and (c), large negative gradients are indicative of a pressure maximum, while large positive gradients are indicative of pressure minima \citep[for an example, see Fig.~1 of][]{Teague_ea_2018a}. We see that the two inner rings for both mm and near-infrared (NIR) emission are centred on pressure maxima, as predicted from grain evolution models \citep{Birnstiel_ea_2012, Pinilla_ea_2012}. The outer ring in scattered light, however, appears at the outer edge of a large pressure minimum centred at $\approx 250$~au. This will be discussed in the following section.

\section{A Perturbed Physical Structure}
\label{sec:pressure_profile}

As the deviations from a smooth rotation curve can be driven by changes in local temperature, density and the height of the emission, it is hard to isolate the main driver. In this section we use a toy model which we perturb in order to reproduce the observed deviations in rotation velocities and infer the underlying pressure profile.

\subsection{The Toy Model}

The model is based on the commonly used prescription using a simple physical structure \citep{Rosenfeld_ea_2013, Williams_Best_2014}, with specific values taken from \citet{Huang_ea_2016}. The total gas surface density is given by \citet{Lynden-Bell_Pringle_1974},

\begin{equation}
\Sigma_{\rm gas}(r) = \Sigma_0 \left( \frac{r}{r_0} \right)^{-\gamma} \exp \left( -\left[\frac{r}{r_0}\right]^{2 - \gamma} \right)
\end{equation}

\noindent where $r_0 = 100~{\rm au}$, $\gamma = 1$ and $\Sigma_0 = 4.95~{\rm g\,cm^{-2}}$ such that the total disk mass is $0.035~M_{\rm sun}$, a factor of 100 times larger than the dust mass used in \citet{Fedele_ea_2018}. This is inflated to a volume density assuming a Gaussian density profile,

\begin{equation}
\rho_{\rm gas}(r,\,z) = \frac{\Sigma_{\rm gas}(r)}{\sqrt{2\pi} H_p(r)} \exp\left( -\frac{z^2}{2H_p(r)^2}\right)
\end{equation}

\noindent with the scale height parametrized as $H_p = 10 \times (r / r_0)^{1.26}~{\rm au}$.

The thermal structure follows the prescription in \citet{Dartois_ea_2003} which smoothly connects two boundary layers, the midplane and the atmosphere, through a trigonometric function,

\begin{equation}
T =
  \begin{cases}
    T_{\rm atm} \quad& z \geq z_q \\
    T_{\rm atm} + \big(T_{\rm mid} - T_{\rm atm}\big) \cos^{2\delta}\left(\frac{z \pi}{2z_q}\right) \quad& z < z_q
  \end{cases} 
\end{equation}

\noindent where $\delta = 2$ and $z_q = 4H_p$. The midplane and atmospheric temperatures are also described by radial power-laws, $T_{\rm mid}(r) = 15.7 \times (r/r_0)^{-0.48}~{\rm K}$ and $T_{\rm atm}(r) = 47.4 \times (r/r_0)^{-0.50}~{\rm K}$, respectively.

Following \citet{Huang_ea_2016}, we consider a homogeneous distribution of CO throughout the disk without taking into account the freeze-out or photodissociation. As the model in \citet{Huang_ea_2016} only considers CO rather than $H_2$, we chose a relative abundance of $x({\rm CO}) = 1.7 \times 10^{-6}$ in order to recover their prescribed column density profiles.

\newpage
\subsection{Perturbations}

We only consider the emission region of the $^{12}$CO which should be narrow in the vertical direction due to the high optical depth of the line. From the model we find that the $^{12}$CO contribution function weighted height, temperature and gas density are well described by power laws over the region of interest, $30~{\rm au} \leq r \leq 320~{\rm au}$:

\begin{equation}
T_{^{12}\rm CO}(r) = 41~{\rm K} \times \left( \frac{r}{100~{\rm au}} \right)^{-0.57},
\end{equation}

\begin{equation}
n_{^{12}\rm CO}(r) = 9.6 \cdot 10^6~{\rm cm^{-3}} \times \left( \frac{r}{100~{\rm au}} \right)^{-2.29},
\label{eq:density}
\end{equation}

\begin{equation}
z_{^{12}\rm CO}(r) = 23~{\rm au} \times \left( \frac{r}{100~{\rm au}} \right)^{1.04}.
\end{equation}

\noindent The subscript $^{12}$CO is to show that these profiles trace the $^{12}$CO emission region and do not necessarily trace a fixed height in the disk.

Taking each of these quantities in turn, we model a perturbation vector as a sum of six Gaussian curves and multiply the power-law describing that property by this to create a perturbed profile \citep[similar to the perturbations used to model the continuum intensity profile][]{Fedele_ea_2018}. Using this perturbed profile and fixing the other two properties, $v_{\rm rot}$ is calculated using Eqn.~\ref{eq:rotation} and compared to the observed $v_{\rm rot}$ profile. Although in reality these three properties are highly coupled, this approach allows us to quantify the extreme cases which are consistent with the data.

\begin{figure}
\centering
\includegraphics[]{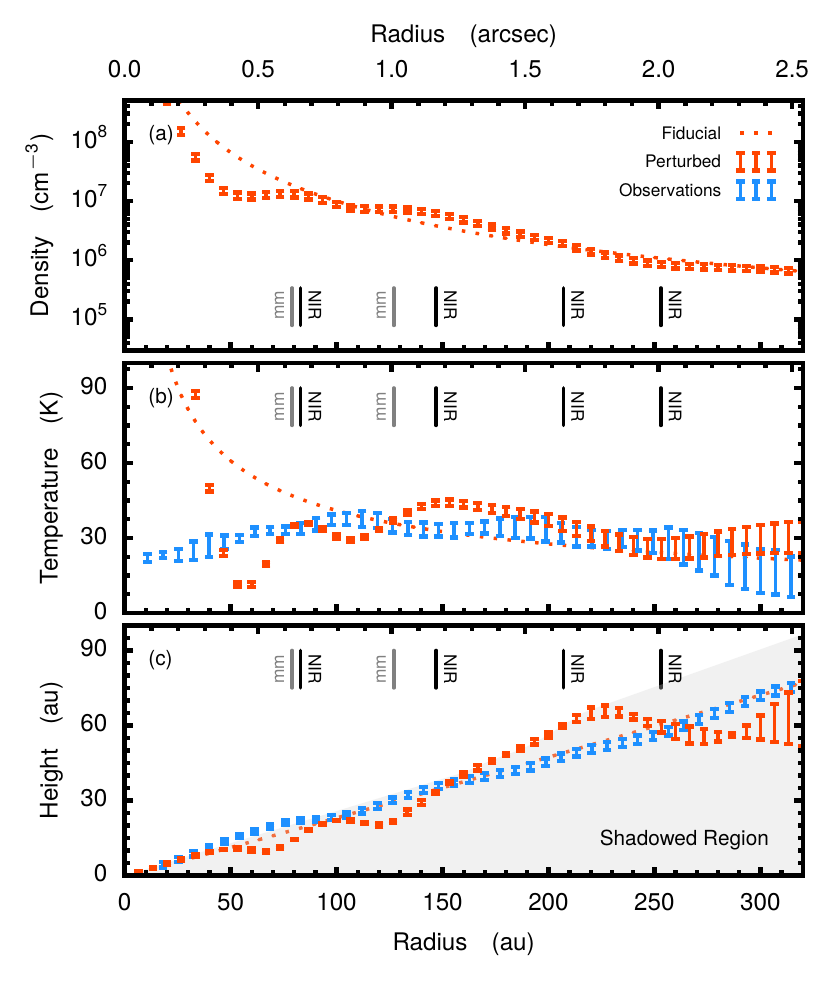}
\caption{Perturbed physical profiles resulting in the observed $v_{\rm rot}$. The red dotted lines show the fiducial models while the red error bars show the 16th to 84th percentile range from 200 draws from the posterior distributions. The blue error bars show the observations: $^{12}$CO peak brightness temperature for the temperature and the inferred emission surface. Note there are no observable constraints on density. Gray and black vertical lines show the location of the rings in mm~continuum and NIR scattered light, respectively. In panel (c) the shaded region shows the shadowed region for the perturbed model showing that many of the NIR scattered rings would be located in shadow. \label{fig:perturbations}}
\end{figure}

The resulting best-fit perturbed profiles are shown in Fig.~\ref{fig:perturbations}. For comparison, the brightness temperature, a proxy of the local temperature for optically thick lines, of the $^{12}$CO emission from the cloud-free region is shown in panel (b) and the derived $^{12}$CO emission surface is shown in panel (c), both with blue error bars. The error bars on the perturbed model represents the scatter of 200 draws from the MCMC fitting.

Decreases in the density are found coincident with the gaps observed in the mm continuum centred at 62 and 103~au \citep{Fedele_ea_2018}. Additionally, a peak is observed at $\approx 150$~au, consistent with the required excess used to explain the ringed CO isotopologue emission \citep{Huang_ea_2016}. 

Large changes in temperature as shown in panel (b) can be ruled out by the line emission as $T_B \leq T_{\rm gas}$ (note however that these temperatures may be underestimated, particularly over the dust rings, due to over subtraction of the dust continuum). Analysis of alternative transitions of $^{12}$CO will help constrain the temperature structure within the inner disk and provide limits for possible changes in temperature, while higher angular resolution will allow for small, local changes to be resolved.
 
Changes in the emission height, shown in panel (c), require larger deviations, particularly between 150 and 250~au, than are allowed from observations. In addition, these perturbations would place the continuum rings centred at local minima in height, shielding the rings from direct irradiation from the star, as shown by the gray shaded region. Such deviations could not be possible given the structure observed in the scattered light \citep{Avenhaus_ea_2018}.

We therefore conclude that the deviations in the rotation velocity are likely driven by change in the radial gas pressure gradient, a combination of both density and temperature.

\subsection{Pressure Profile}

It has been a long standing assumption that grains collect in pressure maxima resulting in environments conducive to grain growth and the beginnings of planet formation \citep{Whipple_1972, Birnstiel_ea_2012, Pinilla_ea_2012}. This is because the velocity of the grains relative to the gas is given by $u_{\rm drift} \propto \partial P \, / \, \partial r$ \citep{Weidenschilling_1977}, then a particle will drift towards pressure maximum. Thus, as grains are predominantly confined within traps this significantly slows radial drift. With constraints on the pressure gradient, we are able to directly test this assumption.

\begin{figure}
\includegraphics[width=\columnwidth]{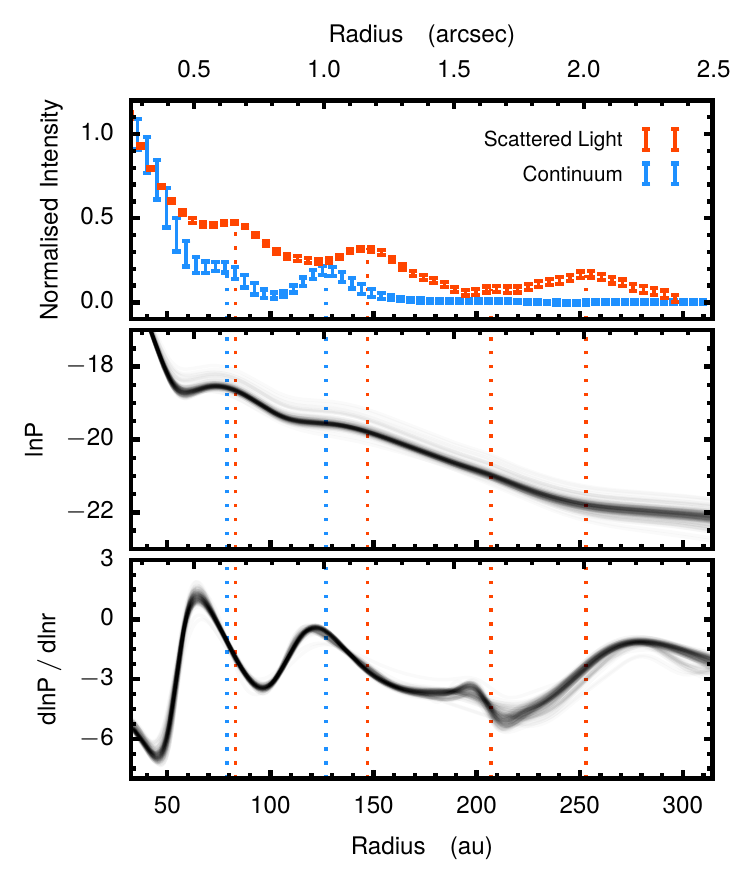}
\caption{Top: Radial profiles of the mm continuum (blue) and the $\micron$ scattered light \citep[red, from][ and scaled by $r^2$ to account for the drop in incident photons]{Avenhaus_ea_2018}. The vertical dashed lines show the location of peaks in these profiles. Middle: samples of the inferred pressure profiles consistent with the measured $v_{\rm rot}$ values. Bottom: The pressure gradient showing the peaks of the scattered light profile align with the pressure maxima. \label{fig:pressure_profile}}
\end{figure}

From the perturbed radial profiles of $T_{^{12}\rm CO}$ and $n_{^{12}\rm CO}$ we can calculate the pressure profile traced by the $^{12}$CO emission. Figure~\ref{fig:pressure_profile} compares this inferred pressure profile and the derivative of its logarithm with the radial continuum emission profile and the $r^2$-scaled scattered light intensity from \citet{Avenhaus_ea_2018}, assuming the southern side of the disk is closest. We see a slight offset in the radial location of the rings in mm and $\micron$ sized grains, with a better match to the pressure maxima with the $\micron$ sized particles.

The absolute scaling of this pressure profile is dependent on the density assumed for the disk model (Eqn.~\ref{eq:density}). Changes in the density structure will result in different amplitude perturbations. Despite this degeneracy, the location of the perturbations will remain constant.

Due to the limited resolution of these observations ($\approx 29$~au for the line emission) the velocity features are not resolved and so will underestimate the true depth. Future, higher resolution observations of the line emission will better constrain the depth of these pressure perturbations and thus their gradient. Such observations will be essential in constraining the level of particle trapping in such pressure maxima.

Only the outer most ring at $\approx 250$~au does not coincide with a pressure maxima, rather with a pressure minima. One possible interpretation of this is that such a drop in pressure will result in decrease of the disk scale height. If this is only a shallow perturbation, such that the far side is not shadowed, then the outer wall of this dip will have a larger angle of incidence for stellar light and thus scatter more effectively from the sub-$\micron$ grains, resulting in a ring despite the lack of a pressure maxima.

\section{Discussion}
\label{sec:discussion}

We have demonstrated that we are able to use gas kinematics to infer the presence of perturbations in the physical structure of AS~209 and infer the radial pressure profile. As these constraints are free of assumptions about the line excitation, they are hugely complimentary to traditional methods aiming to recover emission morphology.

\subsection{Vertical Dependence of Pressure Traps}

Better correlation is found between the pressure maxima and the rings observed in NIR scattered light than the the mm continuum. This suggests that if the rings are due to pressure confinement of the sub-$\micron$ sized grains in the disk atmosphere, there is a vertical dependence in the location of this maxima. This is consistent with the radial offsets found in the location of pressure minima traced at different heights in HD~163296 \citep{Teague_ea_2018a} and with similar features seen in three-dimensional simulations \citep[see, for example, Fig.~4 of][]{Fung_Chiang_2016}.

However, it is unclear whether small particles can as efficiently trapped at higher altitudes than larger, mm-sized particles in the midplane. As the particles become trapped, the rate of collision increases and grain growth is hastened. Larger grains will rapidly settle towards the midplane and drift radially inwards \citep{Dullemond_Dominik_2004}. Two-dimensional simulations combined with accurate vertical profiles for the pressure gradient will be required to properly test this claim.

\subsection{Sources of Perturbations}
\label{sec:discussion:sources_of_perturbations}

To account for the observed deviations in velocity we require at least 3 significant perturbations to the physical structure of the disk at approximately 50~au, 100~au and 250~au. The most attractive scenario for the source of these perturbations is the presence of planets. \citet{Fedele_ea_2018} demonstrated that the continuum emission profile can be explained either by a Saturn mass planet at 95~au and potentially a second planet less than $0.1~M_{\rm Jup}$ at 57~au. The density contrast required to recover the rotation velocities are broadly consistent with thosed used to model the continuum emission profile.

The pressure minima at $\sim 250$~au is less likely to be opened by a planet. Dynamical time scales at these radii are prohibitive for core formation and the formation of planets, although planet formation via gravitational instability is a possibility \citep{Boss_1997}. Recently, \citet{Pinte_ea_2018b} found similar kinematic signatures for a $\sim 2~M_{\rm Jup}$ planet at 260~au in the disk around HD~163296. This suggests that there may be a population of massive, wide separation planets which continuum observations are not sensitive to.

Aside from a planetary origin, other hydrodynamic instabilities have been shown to result in similar perturbations to the disk physical structure. The magneto-rotational instability (MRI) has been shown to drive large gaps in the gas surface density at the outer edge of the dead zone \citep{Flock_ea_2015}. However, estimates of the dead-zone extend only reach out to $\approx 60$~au \citep{Cleeves_ea_2015}, far further in than the observed deviation. Higher resolution observations of molecular line emission will help constrain the local physical properties where perturbations are observed and help distinguish between possible sources.

\subsection{CO Desorption Front}
\label{sec:discussion:co_desorption_front}

\citet{Huang_ea_2016} interpreted the ringed structure observed in C$^{18}$O emission (and tentatively in $^{13}$CO) at 150~au as an enhancement in the local abundance of CO due to a desorption front. The authors argued that the lower opacity of the disk in regions beyond the mm~continuum edge would allow for more efficient non-thermal desorption processes to occur resulting in a local enhancement in CO abundance as hypothesized by \citet{Cleeves_2016}.

As shown in Section~\ref{sec:pressure_profile}, we require enhancements in the H$_2$ gas pressure at 150~au in order to explain the velocity structure, either through an increase in density or temperature. Such changes can also explain the increase in the optically thin line emission without the need for an enhanced in the local CO abundance. However, such changes in the physical structure are likely to also affect the chemistry with an increase in temperature leading to more thermal desorption and higher CO column densities.

\section{Conclusions}
\label{sec:conclusions}

We have extended the method presented in \citet{Teague_ea_2018a} for measuring rotational velocities to allow for non-Gaussian line profiles and for objects with significant azimuthal structure. Application of this method to archival data of AS~209 revealed persistent deviations from a smooth Keplerian profile.

Using a toy model of AS~209, we are able to quantify the deviations required in the temperature, density and height of the emission to match the observed perturbations resulting in deviations of up to 80\%. Future work using models with self-consistent physical structures will be able to disentangle the relative contributions from the density and temperature terms. Comparison of the resulting pressure profiles provides evidence for the pressure trapping of sub~$\micron$ particles in the disk atmosphere, while a radial offset in the ring locations for the mm continuum and the scattered NIR light suggest that the location of the pressure minimum moves radially outwards at higher altitudes in the disk.

A perturbation in the disk structure is inferred at $\approx 250$~au, far beyond the edge of the mm-continuum, resulting in deviations of up to $5\%$ from Keplerian rotation. A planetary origin for this object is unlikely as the dynamical time scales at such large radii make the initial stages of core accretion inefficient.

This work demonstrates the utility of studies of the gas kinematics and the ability to provide unique constraints for the interpretation of high angular resolution continuum observations.

\acknowledgments{We thank the referee for their helpful comments which have made the presentation of this method much clearer. This paper makes use of the following ALMA data: JAO.ALMA\#2015.1.00486.S. ALMA is a partnership of European Southern Observatory (ESO) (representing its member states), National Science Foundation (USA), and National Institutes of Natural Sciences (Japan), together with National Research Council (Canada), National Science Council and Academia Sinica Institute of Astronomy and Astrophysics (Taiwan), and Korea Astronomy and Space Science Institute (Korea), in cooperation with Chile. The Joint ALMA Observatory is operated by ESO, Associated Universities, Inc/National Radio Astronomy Observatory (NRAO), and National Astronomical Observatory of Japan. The National Radio Astronomy Observatory is a facility of the National Science Foundation operated under cooperative agreement by Associated Universities, Inc. This project has received funding from the European Research Council (ERC) under the European Union’s Horizon 2020 research and innovation programme under grant agreement No 714769 and was supported by funding from NSF grants AST-1514670 and NASA NNX16AB48G. R.T. would like to thank Bertram Bitsch, Kees Dullemond and Mario Flock for insightful discussions and Henning Avenhaus for sharing the scattered light data.}

\software{bettermoments \citep[][]{Teague_Foreman-Mackey_2018}, 
	      CASA \citep[v4.2 \& v5.2, ][]{McMullin_ea_2007}, 
          celerite \citep[][]{Foreman-Mackey_ea_2017},
          eddy \citep[][]{eddy},
          emcee \citep[][]{Foreman-Mackey_ea_2013}, 
          matplotlib \citep[][]{Hunter_2007}, 
          numpy \citep[][]{vanderWalt_ea_2011}, 
          scipy \citep[][]{Jones_ea_2001}}

\appendix

\section{Recovering the Rotation Velocity}
\label{sec:app:rotation_velocity}

In this Appendix we demonstrate the robustness of the derived $v_{\rm rot}$ and quantify the precision which can be achieved with this method. The code used to calcaulted $v_{\rm rot}$ and the model spectra can be found at \url{https://github.com/richteague/eddy}. Version 1.0 of the code used for this paper can be obtained from \url{https://doi.org/10.5281/zenodo.1440051}.

\subsection{General Properties}
\label{sec:app:general_properties}

We first consider the case of well behaved data: intrinsic Gaussian profiles with Gaussian noise. To model the $^{12}$CO emission we generated 20,\,000 sets of model data. Each sample contained a random number of spectra, $N \in [6, \, 60]$, linearly spaced across the $2 \pi$ azimuth. This range encompasses the expected number of independent beams for disk observed with ALMA at $\sim 0.1\arcsec$ resolution: $N \approx 2 \pi r \, / \, \theta_{\rm beam}$.

The underlying profile was assumed to be Gaussian described by  $T_{\rm B} \in [5, \, 40]~{\rm K}$, $\Delta V \in [100, \, 400]~{\rm m\,s^{-1}}$ and $v_{\rm rot} \in [0.5, \, 3.5]~{\rm km\,s^{-1}}$. Each was then corrupted by Gaussian noise to achieve a ${\rm SNR} \in [2, \, 20]$. These values were chosen to represent typical line properties observed in protoplanetary disks. They were calculated on a velocity axis with a resolution of $160~{\rm m\,s^{-1}}$ meaning that the FWHM of the line was sampled between roughly 1 and 4 times.

\begin{figure}
\centering
\includegraphics[]{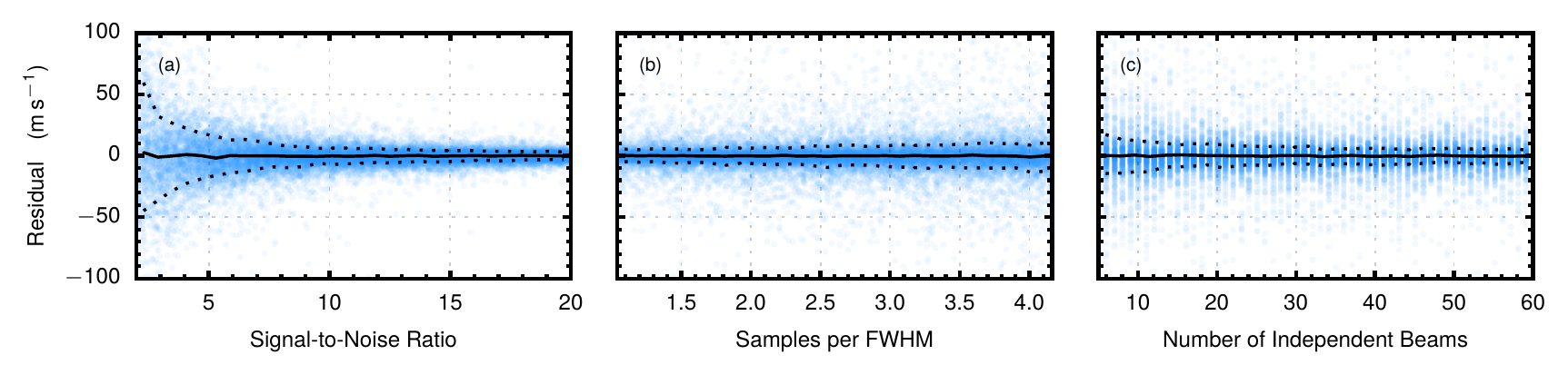}
\caption{Accuracy and preicision achieved for the method described in Section~\ref{sec:methodology} while varying the signal-to-noise of the spectra, the linewidth of the line and the number of spectra used. The blue points each represent a random sample, while the solid like shows the median of the distribution and the dotted show the 16th and 84th percentiles (equivalent to one standard deviation for a Gaussian distribution). \label{fig:app:precision_GP}}
\end{figure}

For each sample of lines, we inferred $v_{\rm rot}$ following the method described in Section~\ref{sec:methodology}. The difference from the true value are shown in Fig.~\ref{fig:app:precision_GP}. As expected, the accuracy achieved increases with the SNR of the data and the number of lines used as both of these increase the SNR of the stacked spectra. Marginalizing over all intrinsic line properties we can model the accuracy of this method via the power-law it to the 16th to 84th percentile range of residuals (shown by the dotted lines in Fig.~\ref{fig:app:precision_GP}) as,

\begin{equation}
{\rm Accuracy} = 44.4 \pm 0.1~{\rm m\,s^{-1}} \times \left( 
\sqrt{\frac{2 \pi r}{\theta_{\rm beam}}} \cdot \frac{{\rm SNR}}{10} \right)^{-1.12 \pm 0.01},
\end{equation}

\noindent showing that with SNRs readily achievable by ALMA (due to both the overall sensitivity and the small beam sizes achieved in order to better sample the annulus), accuracies of a few meters per second are possible. A measure of the precision of the results can be estimated by the width of the posterior distribution for $v_{\rm rot}$ for each case. This is well fit with the profile,

\begin{equation}
{\rm Precision} = 23.9 \pm 0.6~{\rm m\,s^{-1}} \times \left( 
\sqrt{\frac{2 \pi r}{\theta_{\rm beam}}} \cdot \frac{{\rm SNR}}{10} \right)^{-1.39 \pm 0.04},
\end{equation}

\noindent which is roughly a factor of two larger than the accuracy. Thus the assumed $3\sigma$ uncertainties quoted in this work \citep[and][]{Teague_ea_2018a}, are consistent with the true $v_{\rm rot}$ value.

\subsection{Azimuthal Asymmetry}
\label{sec:app:azimuthal_asymmetry}

For the case of AS~209 there is strong azimuthal asymmetry due to the cloud absorption. In this section we demonstrate how such azimuthal structure in either $T_{\rm B}$ or $\Delta V$ affects the inferred $v_{\rm rot}$. Similar to the previous examples we generate model spectra, however reduce the parameter space by considering only $\Delta V = 300~{\rm m\,s}^{-1}$, $N = 20$ and ${\rm SNR} = \{5,\, 10,\, 15\}$. We then include a periodic perturbation in $\Delta V$, $T_{\rm B}$ or both parameters, parameterised as

\begin{equation}
\delta = 1 + \delta_0 \cdot \sin \left( \frac{\theta + \chi}{f} \right)
\label{eq:app:azimuthal_deviation}
\end{equation}

\noindent where $\delta_0$ controls the strength of the deviation, $\chi \in [-\pi, \, \pi)$ is a random number to offset the deviation and $f$ is an integer frequency. For this Appendix we only consider $f = \{1,\, 2\}$ and $\delta_0 \in [0,\, 0.5]$. For the case of AS~209, the cloud contamination leads to a $\delta_0 \approx 0.3$.

\begin{figure}
\centering
\includegraphics[]{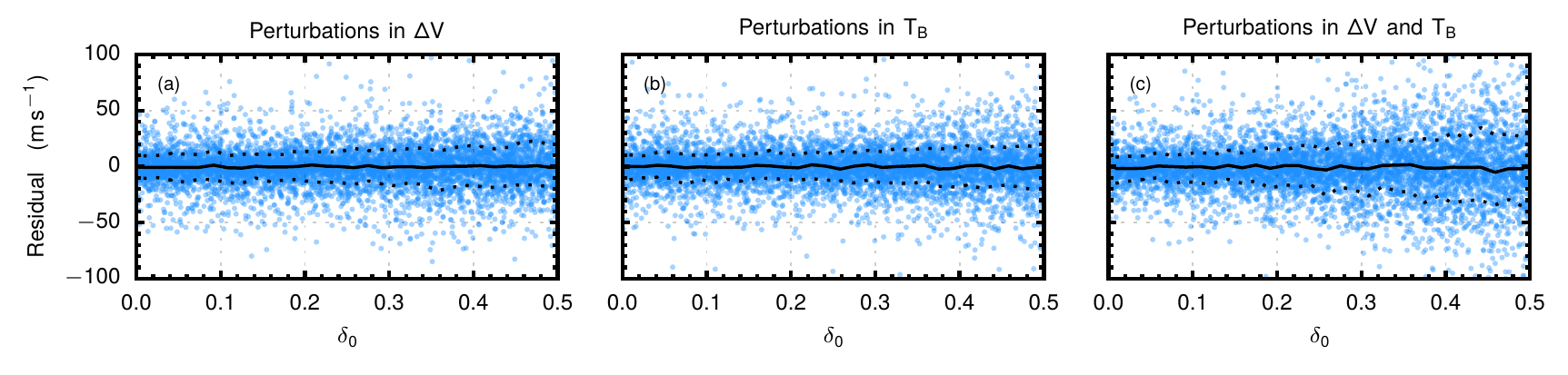}
\caption{Accuracy of the Gaussian Process method when azimuthal structure is considered. Each call randomly select $\{\delta_0,\, \chi,\,f\}$, as described in Eqn.~\ref{eq:app:azimuthal_deviation}. The left panel shows when the deviation is only applied to the linewidth, the center panel to the line peak and the right panel, both parameters. \label{fig:app:precision_GP_azimuthal}}
\end{figure}

The results are shown in Fig.~\ref{fig:app:precision_GP_azimuthal} where each panel shows 1,\,100 samples. The dotted lines show the 16th and 84th percentiles of the distribution and the solid line shows the median. These show that accuracy is not strongly affected by the inclusion of azimuthal structure. This suggests that the method is able to robustly recover the an accurate measure of the line to a an accuracy of $\lesssim 20~{\rm m\,s^{-1}}$ even when both $\Delta V$ and $T_{\rm B}$ have perturbations of up to 30\%.

\subsection{Spatially Resolved 3D Structure}
\label{sec:app:3d_structure}

\begin{figure}
\centering
\includegraphics[]{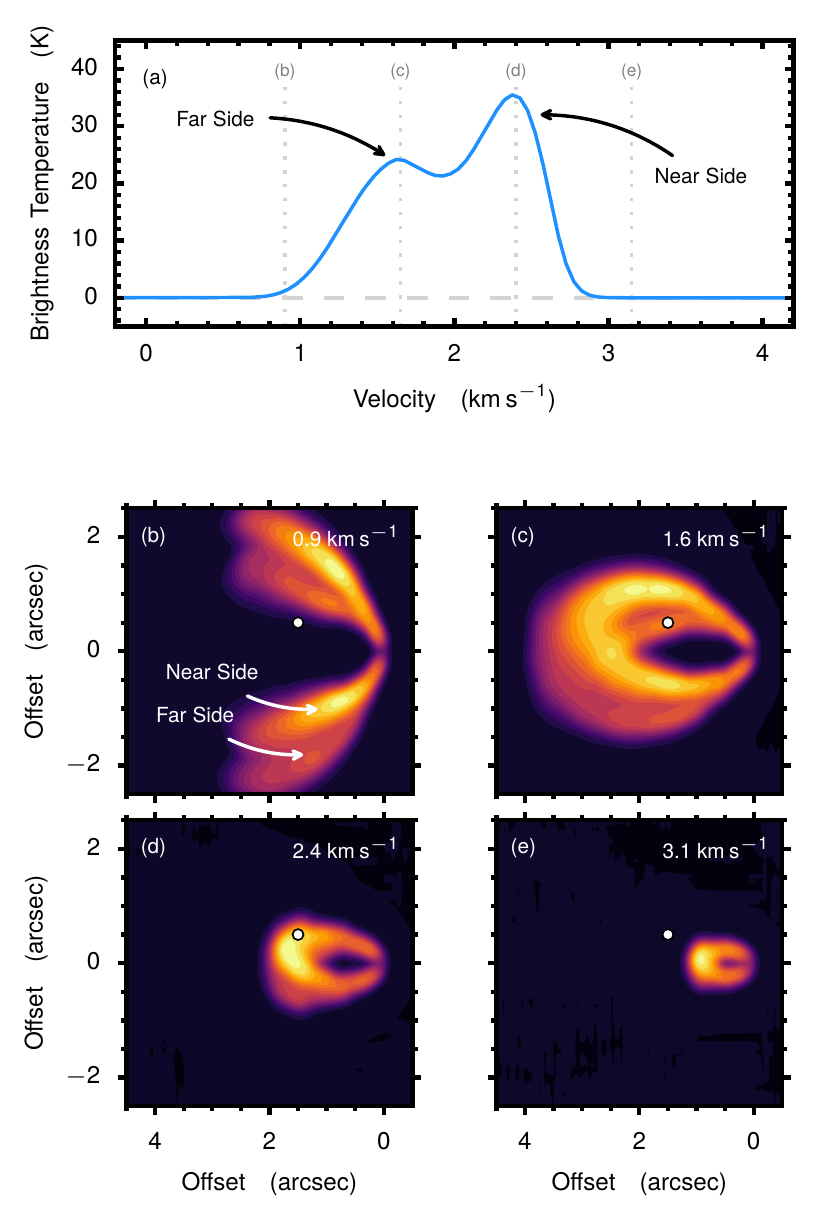}
\caption{Demonstrating the origin of the double Gaussian line profiles for high spatial resolution data. The top panel shows the spectrum extracted at $(1.5\arcsec,\,0.5\arcsec)$, shown by the white dot in the channel maps. The velocities of the channel maps are shown in the top right of each panel and as a vertical dotted line in the spectrum. For a given position we find two Gaussian components due to the near and far sides of the disk with the near side being brighter. \label{fig:double_gaussian}}
\end{figure}

With ALMA now able to routinely spatially resolve the near and far side of the disk for molecules with high emission surfaces \citep{deGregorio-Monsalvo_ea_2013, Rosenfeld_ea_2013}, emission from both the top and bottom sides of the disk will be visible along a line of sight. This results in two components rather than a single component which could potentially cause problems as demonstrated in Fig.~\ref{fig:double_gaussian}.

To demonstrate the robustness of this technique against such contamination, we consider a simple model set-up. Each line of sight will be the combination of a main Gaussian component from the near side of the disk and a slightly offset secondary peak from the rear side of the disk. The secondary peak will have a smaller amplitude as this will be tracing the snow-surface of the far side of the disk \citep{Pinte_ea_2018a}, thus $T_B \approx 21$~K.

To calculate the offset of the secondary peak we consider a disk with a given stellar mass, $M_{\star}$, inclination, $i$, and emission surface described by $z = z_0 \times r^{\phi}$, where $z_0 = 0.3$ and $\phi = 1.25$. For a given radius we first calculate the mapping of disk coordinates to sky coordinates via \citep[see also][]{Rosenfeld_ea_2013},

\begin{equation}
\left(\begin{array}{c} x_{\rm sky} \\ y_{\rm sky} \end{array}\right) = \left(\begin{array}{c} x_{\rm disk} \\ \big(y_{\rm disk} - z_{\rm disk} \, / \, \sin(i) \big) \cdot \cos(i) \end{array}\right).
\end{equation}

\noindent We then calculate where these sky coordinates intercept the far side of the disk. For the emission surface of the far side of the disk we consider a smaller aspect ratio of $z \,/\, r = 0.1$ as this emission will arise from the snow surface, as discussed previously, and thus be tracing a deeper (closer to the midplane) region. This results in two sets of coordinates: $(x_{\rm disk}, \, y_{\rm disk}, \, z_{\rm disk})$ for the front side of the disk and the same for the far side of the disk. The front side coordinates will describe a ring of radius $r_{\rm disk}$ with height $z_{\rm disk}$ while the rear side will sample a range of $r_{\rm disk}$ values and thus different $z_{\rm disk}$ and $v_{\rm rot}$ values.

\begin{figure}
\centering
\includegraphics[]{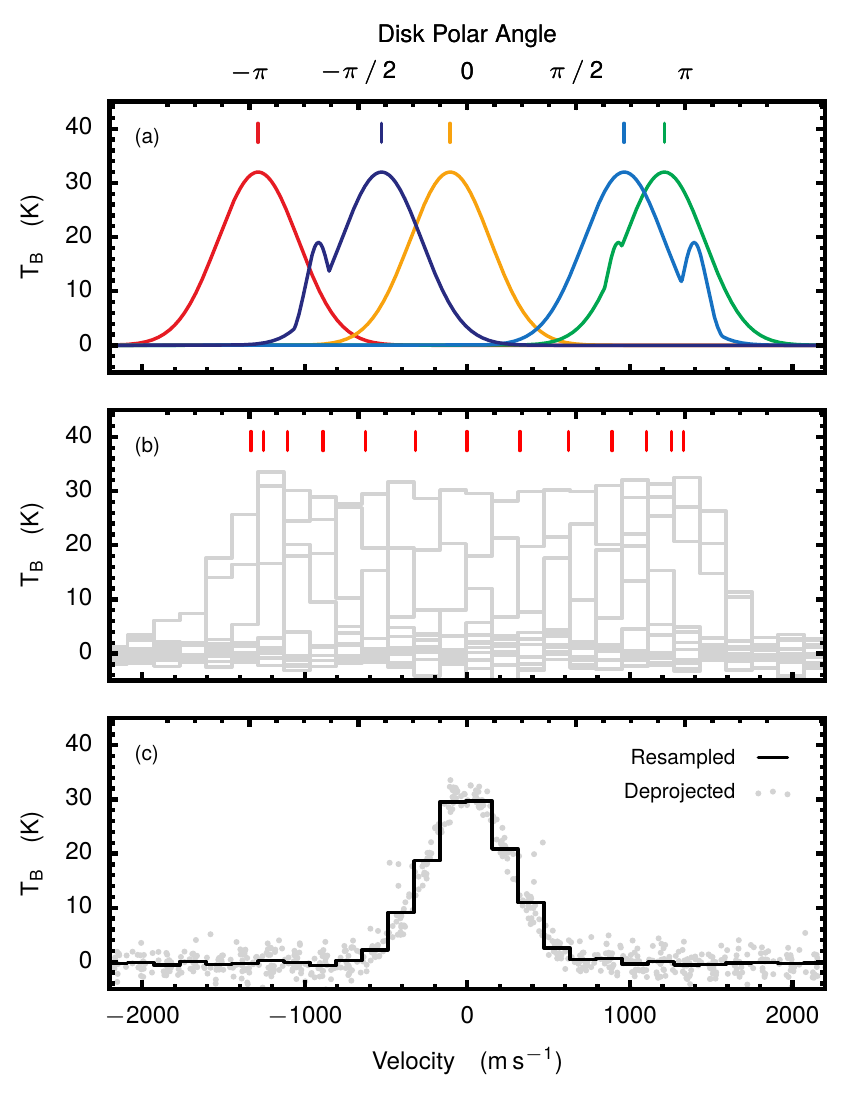}
\caption{Examples for a flared disk. The top panel shows example spectra including the contamination from the rear side of the disk. The middle panel shows example spectra for an AS~209 like disk with parameters matching those described in Section~\ref{sec:as209:observations}. The red lines indicate the line centres. The bottom panel shows the results of the deprojection showing that the contamination from the far side does not significantly perturb the final stacked spectra. \label{fig:app:flared_disk_example}}
\end{figure}

Figure~\ref{fig:app:flared_disk_example}a shows example spectra for disk matching the parameters of AS~209. The contamination from the rear side of the disk is seen as small components offset from the main line centre with the largest deviations in the regions between the major ($\theta = 0$) and minor ($\theta = \pm \pi$) axes.

In Fig.~\ref{fig:app:flared_disk_example}b we show an annulus of spectra including the appropriate noise ($\sigma_{\rm rms} \approx 2~{\rm K}$) and channel width ($\Delta V_{\rm chan} = 160~{\rm m\,s^{-1}}$). The red lines show the line centers before deprojection. Figure~\ref{fig:app:flared_disk_example}b shows the deprojected spectra (in gray dots) and the stacked profile as a solid line. As the rear side components are varying in their offset for each line, they do not stack coherently and are thus lost in the noise, allowing for a good fit of $v_{\rm rot}$ to be found even in this scenario, achieving an accuracy of $< 0.4\%$.

In the case of very high signal-to-noise data with very small beam sizes, the far side components will become more apparent. This can be circumvented using the initial method of minimizing the width of a Gaussian line profile as this prior will be less sensitive to the contamination.

\end{document}